\documentclass[aps,10pt,pre,twocolumn,superscriptaddress]{revtex4-1}

\usepackage{graphicx,color,bm,natbib,placeins,times}
\usepackage{amsmath,amssymb}
\usepackage{enumitem} 

\begin{document}
\title{Incoherence-Mediated Remote Synchronization}

\author{Liyue Zhang}
\affiliation{Center for Information Photonics and Communications, Southwest Jiaotong University, Chengdu, Sichuan 610031, China}
\affiliation{Department of Physics and Astronomy, Northwestern University, Evanston, IL 60208, USA}

\author{Adilson E. Motter}
\affiliation{Department of Physics and Astronomy, Northwestern University, Evanston, IL 60208, USA}
\affiliation{Northwestern Institute on Complex Systems, Northwestern University, Evanston, IL 60208, USA}

\author{Takashi Nishikawa}
\email{t-nishikawa@northwestern.edu}
\affiliation{Department of Physics and Astronomy, Northwestern University, Evanston, IL 60208, USA}
\affiliation{Northwestern Institute on Complex Systems, Northwestern University, Evanston, IL 60208, USA}

\begin{abstract}
In previously identified forms of remote synchronization between two nodes, the intermediate portion of the network connecting the two nodes is not synchronized with them but generally exhibits some coherent dynamics. Here we report on a network phenomenon we call {\it incoherence-mediated remote synchronization} (IMRS), in which two non-contiguous parts of the network are identically synchronized while the dynamics of the intermediate part is statistically and information-theoretically incoherent. We identify mirror symmetry in the network structure as a mechanism allowing for such behavior, and show that IMRS is robust against dynamical noise as well as against parameter changes. IMRS may underlie neuronal information processing and potentially lead to network solutions for encryption key distribution and secure communication.
\end{abstract}

%

\onecolumngrid\hfill 
{\small Published in Phys. Rev. Lett. {\bf 118}, 174102 (2017)}
\bigskip\twocolumngrid

\maketitle 

Communication, broadly defined as information exchange between different parts of a system, is a fundamental process through which collective dynamics arises in complex systems.
Network synchronization~\cite{Pikovsky2003}, whether it is complete 
synchrony~\cite{syn5} 
or a more general form of synchronization~\cite{Kaneko:1990,Stewart:2003,cluster2,cluster1,pecora}, is a primary example of such dynamics and is thought to be largely driven by node-to-node communication.
However, it has recently been shown that so-called remote 
synchronization~\cite{Winful:1990,Fischer:2006,Sande2008,Soriano2012,Bergner:2012,Viriyopase:2012,remote2,Gambuzza:2013} is possible: two distant nodes (or groups of nodes) can synchronize even when the intermediate nodes are not synchronized with them.
In this form of synchronization, the dynamics of different intermediate nodes generally show some level of coherence with each other, exhibiting, e.g., generalized synchronization or delay synchronization.

\begin{figure}[t] 
\includegraphics[width=0.92\columnwidth]{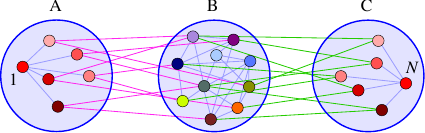}
\caption{
Remote synchronization between node groups A and C mediated by incoherence in group B.
The colors of the nodes schematically represent their states, indicating that nodes $1$ and $N$ are identically synchronized, while the dynamics of the nodes in B are incoherent.
}\label{fig0}
\end{figure}

In contrast, in this Letter we consider a dynamical state of a network that we shall call {\it incoherence-mediated remote synchronization} (IMRS).
The $N$ nodes of the network are organized into three non-empty groups, A, B, and C, where A is connected with B, and B is connected with C, but A and C are not directly connected (as illustrated in Fig.~\ref{fig0}).
We assume that group B has at least two nodes, and that the nodes and links within each group form a connected subnetwork.
IMRS is then characterized by (1) a node from group A (denoted node $1$) and a node from C (denoted node $N$) that are identically synchronized (rather than in weaker forms such as phase and generalized synchronization), and (2) the dynamics of the nodes in the intermediate group B that are statistically incoherent with each other.
IMRS combines the properties of remote synchronization mentioned above with those of chimera 
states~\cite{Kuramoto:2002,Abrams:2004,pecora1,Tinsley:2012,Panaggio:2015}, 
which are characterized by the coexistence of both coherent and incoherent dynamics in different parts of the network.
Here, however, we lift the assumption of uniform network typically made in studying chimera states, and instead ask the following fundamental question: under what conditions can IMRS be observed?
In particular, what types of network structure allow for this behavior?
Below we answer these questions by mapping them to the problem of cluster synchronization and using a powerful tool for studying network symmetry based on computational group theory~\cite{pecora}.
Moreover, we show that the incoherent dynamics of group B is typically also incoherent relative to the dynamics of node $1$ (and $N$).
This suggests applications of IMRS to new forms of secure communication technologies~\cite{Cuomo:1993,Argyris:2005} or new schemes for secure  generation and distribution of encryption keys~\cite{hayato}.

We consider a general class of networks of $N$ coupled identical dynamical units, whose time evolution is governed by
\begin{equation}\label{eqn:system}
\dot{\mathbf{x}}_i = \mathbf{F}(\mathbf{x}_i) + \sigma \sum_{j=1}^N A_{ij} \mathbf{H}(\mathbf{x}_j),
\end{equation}
where $\mathbf{x}_i(t)$ is the state of the $i$th unit at time $t$, the equation $\dot{\mathbf{x}} = \mathbf{F}(\mathbf{x})$ describes the dynamics of an isolated node, $\sigma$ is the overall coupling strength, $A = (A_{ij})_{1\le i,j\le N}$ is the coupling matrix representing an undirected unweighted network topology of the type illustrated in Fig.~\ref{fig0}, and $\mathbf{H}(\mathbf{x})$ is a function determining the output signal from a node.
Within this framework,
we formulate a set of three conditions for IMRS to be observed:
\begin{enumerate}[label=(\roman*)]
\item There exists a state in which $\mathbf{x}_1(t)=\mathbf{x}_N(t)$ for $\forall t$.
\item The state of synchronization between nodes $1$ and $N$ in condition (i) is stable.
\item $\{\mathbf{x}_i(t)\}$ and $\{\mathbf{x}_j(t)\}$ are not synchronized for all node pairs and are statistically incoherent for most pairs in B.
\end{enumerate}
(Recall that nodes $1$ and $N$ are from groups A and C, respectively.)

Although condition (i) is dynamical in nature, a network-structural condition implying condition (i) can be expressed solely in term of the symmetry of the network.
The network symmetry is represented by the (mathematical) group of node permutations under which the network structure is invariant (or, equivalently, the group consisting of the corresponding permutation matrices that commute with the adjacency matrix $A$).
A cluster of synchronized nodes can be identified as an orbit of this group, defined as a set of nodes in which each node can be mapped to any other nodes in the set by some permutation in the group.
From the invariance of Eq.~\eqref{eqn:system} under these permutations, it follows that there is a synchronous state in which all nodes in each orbit (of the group) have identical dynamics, forming $K$ clusters: $\{\mathbf{s}_k(t)\}_{1\le k\le K}$, where $\mathbf{x}_i(t)=\mathbf{s}_k(t)$ for all $t$ if node $i$ belongs to cluster $C_k$.
Note that $\mathbf{s}_k(t)$ can be different for different $k$ as long as they satisfy the equations obtained by substituting $\mathbf{x}_i(t)=\mathbf{s}_k(t)$ into Eq.~\eqref{eqn:system}.
Formulating IMRS as such a state, we see that condition~(i) above is equivalent to the existence of an orbit that intersects with both A and C.
We denote this cluster by $C_1$, from which we choose one node in A as node $1$, and one node in C as node $N$.

The synchronization stability condition (ii) is verified for a given network structure using the method in Ref.~\cite{pecora}.
We first identify clusters $C_k$ in the network using computational group theory.
We then compute $\lambda_{C_1}$, the maximum transverse Lyapunov exponent associated with the modes of perturbation that destroys the synchronization of cluster $C_1$ (some of which destroy the synchronization between nodes $1$ and $N$).
Thus, condition~(ii) can be formulated as $\lambda_{C_1}<0$.

The statistical coherence in condition (iii) is measured by cross correlation and mutual information, accounting for possible coherence with a time lag $\Delta t$.
We use $C_{i,j}$ to denote the absolute value of the Pearson correlation coefficient between $\mathbf{x}_i(t)$ and $\mathbf{x}_j(t+\Delta t)$ over a range of $t$, maximized over a range of $\Delta t$~\cite{timedelay1}.
Likewise, we use $I_{i,j}$ to denote the mutual information between $\mathbf{x}_i(t)$ and $\mathbf{x}_j(t+\Delta t)$ over $t$, maximized over $\Delta t$~\cite{timedelay1,mi1}.
Thus, condition~(iii) would be satisfied if $C_{i,j}$ and $I_{i,j}$ are both small for most pairs $i$ and $j$ in B, and $C_{i,j}\neq 1$ for $\forall i,j$ (indicating no identical synchronization).
We choose chaotic node dynamics for higher likelihood of having incoherence in B, and we further ensure that the dynamics of $\mathbf{s}_k(t)$ is chaotic.
This condition is equivalent to $\lambda>0$, where $\lambda$ is the maximum Lyapunov exponent parallel to the synchronization manifold (associated with perturbations that do not destroy synchronization of any cluster $C_k$).

Condition~(iii) is also intimately related to network symmetry; it
requires that each cluster in B contain only one node.
What characterizes the structure of networks that satisfy both this requirement and condition~(i)?
Based on our numerical verification for $N\le 8$ nodes,
we conjecture that any such network has a mirror symmetry (possibly after regrouping the nodes): groups A and C are ``mirror images'' of each other (although no symmetries are needed inside group B), as illustrated in Fig.~\ref{fig0}.
More precisely, the network structure is invariant under a node permutation that serves the role of a ``reflection'' and maps each node in A to a unique node in C, but does not move any nodes in B.
In particular, this implies that each node in B that connects to A must connect to C in exactly the same way.
It also implies that all nontrivial clusters (i.e., those of size $>1$), which we denote $C_1,\dots,C_{K'}$ (after appropriate re-indexing), span both A and C in a symmetrical way (involving the same number of nodes from each group) and collectively cover all nodes in A and C.
This means that the corresponding network dynamics is also mirror symmetric: each node in A is identically synchronized with its counterpart in C (possibly showing different dynamics for different node pairs).
In particular, we have identical synchronization between nodes $1$ and $N$ (both belonging to $C_1$).
Moreover, the clusters $C_1,\dots,C_{K'}$ are all intertwined with each other, i.e., synchronization of these clusters must be either all stable or all unstable.
A group-theoretical origin of this behavior is argued to be the property that any network-invariant permutation that rearranges the nodes in one cluster must also rearrange the nodes in each of the other clusters~\cite{pecora}, which we conjecture is guaranteed by the mirror symmetry.
Conversely, if a network with the three-group structure of Fig.~\ref{fig0} has a mirror symmetry, then nodes $1$ and $N$ (in A and C, respectively) are guaranteed to be part of a synchronized cluster.
Note that the mirror symmetry alone does not impose any condition on the link configuration within B, and hence the clusters in B can in principle be of size $>1$ [which would violate condition (iii)].

\begin{figure}[t]
\centering
\includegraphics[width=\columnwidth]{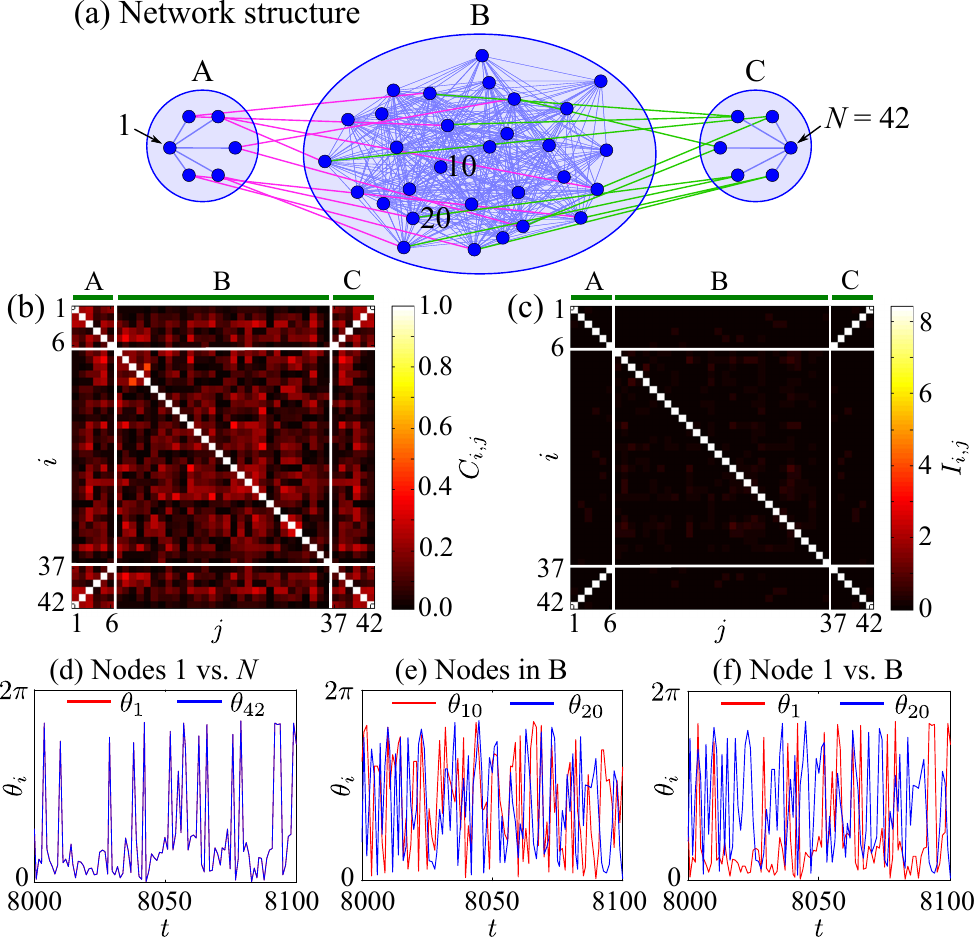}\vspace{-3mm}
\caption{
Network exhibiting IMRS.
(a) Mirror-symmetric structure of the network, generated with $n_\text{A}=6$, $n_\text{B}=30$, $n'_\text{B}=2$, and $p=0.8$.
(b) Pairwise cross correlation $C_{i,j}$.
(c) Pairwise mutual information $I_{i,j}$.
(d)--(f) Phase variable $\theta_i$ as a function of $t$.
In (d), only the blue curve is clearly visible because the two curves overlap. 
The calculations in (b)--(f) are based on iterating Eq.~\eqref{discrete} with $\beta=1.5$ and $\sigma=1.5$.
}\label{fig3}
\end{figure}

To systematically search for IMRS, we propose the following general recipe for designing a system:
1) construct a network structure that has a mirror symmetry and satisfies the size-one cluster requirement in B;
2) select chaotic node dynamics; 3) find system parameters for which the synchronization between nodes $1$ and $N$ is stable (i.e., $\lambda_{C_1}<0$) and the dynamics of $\mathbf{s}_k(t)$ is chaotic (i.e., $\lambda>0$); and 4) verify incoherence in B (i.e., small $C_{i,j}$ and $I_{i,j}$).
As an example algorithm for generating networks for step $1$ above, we use the following procedure (for which we provide software; see Supplemental Material~\cite{sm}).
Given $n_\text{A}$, $n_\text{B}$, and $n_\text{C}$ ($=n_\text{A}$) nodes in A, B, and C, respectively, we first connect each pair of nodes in B with probability $p$.
Next, we connect node $1$ to all the other nodes in A and node $N$ to all the other nodes in C.
The nodes in A other than node $1$ are then paired up with the nodes in C other than node $N$.
Finally, for each of these node pairs, we choose $n'_\text{B}$ nodes randomly from B and connect each of these nodes to the node pair.
An example network constructed by this procedure is shown in Fig.~\ref{fig3}(a).
The probability of having a cluster of size $>1$ in B can be kept small by making the size of B large enough.
Here we generate networks with $n_\text{B}\ge 10$ and use only those with no cluster of size $>1$ in B.

\begin{figure}[t] 
\includegraphics[width=1.0\columnwidth]{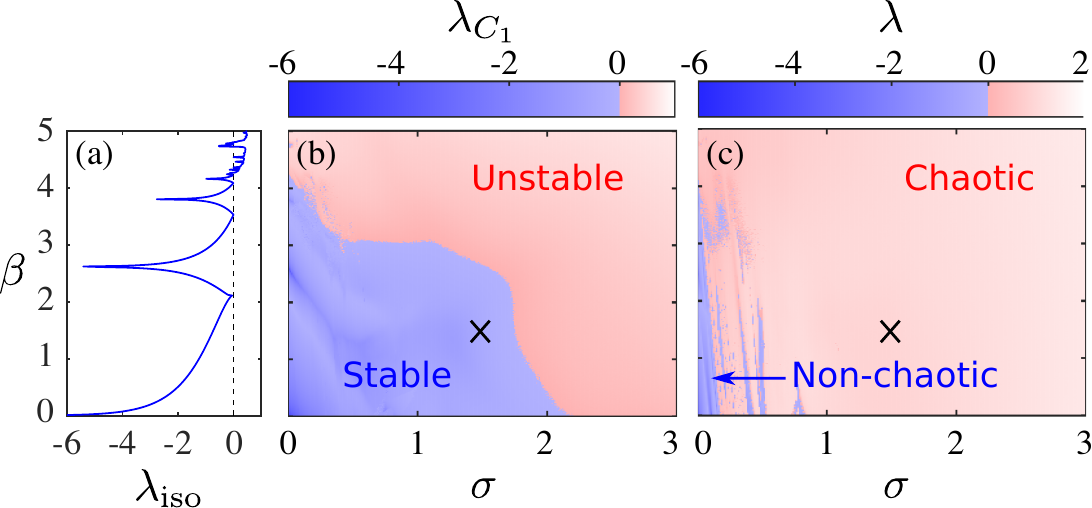}\vspace{-2mm}
\caption{
Characterizing the network dynamics.
(a) Lyapunov exponent $\lambda_\text{iso}$ of the isolated node dynamics as a function of self-feedback strength $\beta$.
(b) Synchronization stability $\lambda_{C_1}$ of cluster $C_1$ (and thus between nodes $1$ and $N$) as a function of $\beta$ and coupling strength $\sigma$.
(c) Lyapunov exponent $\lambda$ measuring the instability parallel to the synchronization manifold as a function of $\beta$ and $\sigma$.
The exponents $\lambda_{C_1}$ and $\lambda$ are averaged over $10$ network realizations and $10$ initial conditions.
\label{fig2}} 
\end{figure}

As an example dynamics for the network leading to IMRS, we use coupled maps that model the electro-optic experimental system~\cite{pecora1}, although we note that continuous-time systems also exhibit IMRS (see Supplemental Material~\cite{sm}, Sec.~\ref{sec:lorenz}).
The system dynamics is governed by  
\begin{equation}
\theta_i(t+1)=\biggl[\beta\,\emph{I}\bigl(\theta_i(t)\bigr)+\sigma\sum_{j=1}^N A_{ij}\emph{I}\bigl(\theta_j(t)\bigr)+\delta\biggr]\textrm{mod} \ 2\pi,
\label{discrete}
\end{equation}
where $\theta_i(t)$ is the phase shift in time step $t$ for the $i$th component of the spatial light modulator array used in the experiment, $\beta$ is the strength of self-feedback coupling for the array components, and the offset $\delta$ is introduced to suppress the trivial solution, $\theta_i(t)\equiv0$.  We set $\delta=0.525$ for all computations for this system. 
The intensity of light is related to the phase shift $\theta$ through the nonlinear function $I(\theta):=[1-\cos(\theta)]/2$. 
The dynamics of an isolated oscillator has a globally stable fixed point for small $\beta$, which, through a sequence of period-doubling bifurcations, becomes chaotic for larger values of $\beta$ [see Fig.~\ref{fig2}(a)].

As shown in Fig.~\ref{fig2}(b), we find that networks generated by the procedure described above can achieve $\lambda_{C_1}<0$ (i.e., stable synchronization between nodes $1$ and $N$)
when $\beta$ and $\sigma$ are both relatively small.
These networks all have a mirror symmetry by construction, and they
satisfy both conditions~(i) and (ii).
Figure~\ref{fig2}(c) shows that, even when we start with oscillators that are not chaotic in isolation [$\beta\lesssim 4$, see Fig.~\ref{fig2}(a)], the dynamics of the clusters $\mathbf{s}_k(t)$ becomes chaotic (i.e., $\lambda>0$) as the coupling strength $\sigma$ is increased.
We thus see that there is a wide range of parameters $\beta$ and $\sigma$ for which the network realizes stable chaotic synchronization.
To check condition~(iii), we compute $C_{i,j}$ and $I_{i,j}$ over time steps $10^4\leq t\leq4\times 10^4$ and time delay $-50\le\Delta t \le 50$ for $\beta=1.5$ and $\sigma=1.5$ [black crosses in Figs.~\ref{fig2}(b) and \ref{fig2}(c)].
The results, shown in Figs.~\ref{fig3}(b) and \ref{fig3}(c), verify that condition~(iii) is indeed satisfied.
The corresponding system dynamics is illustrated by the time plots in Figs.~\ref{fig3}(d)--(f).
Thus, the network exhibits IMRS for these specific parameters.
Moreover, Figs.~\ref{fig3}(b) and ~\ref{fig3}(c) clearly show that the dynamics of the nodes in B is also incoherent relative to nodes $1$ and $N$.
While Eq.~\eqref{discrete} is a discrete-time analog of Eq.~\eqref{eqn:system}, we expect IMRS to be observed for a range of different node dynamics, including both discrete-time and continuous-time dynamics, as well as for many mirror-symmetric network topologies not necessarily generated by the procedure described above.

\begin{figure*}[t]
\centering
\includegraphics[width=\textwidth]{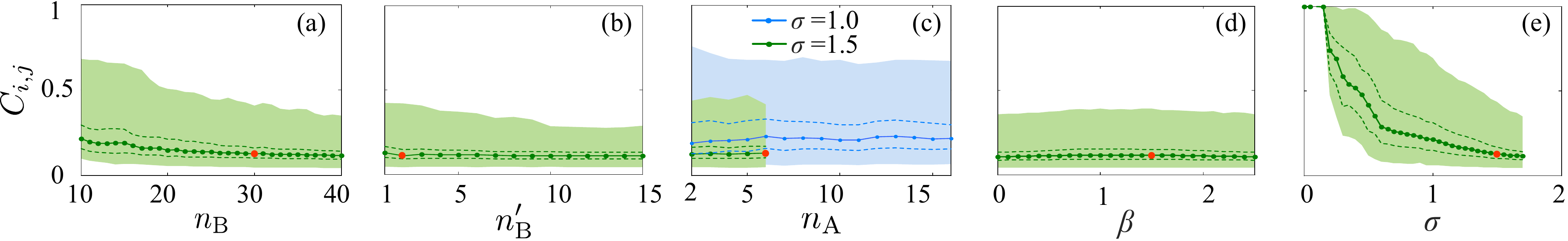}\vspace{-4mm}
\caption{
Influence of system parameters on IMRS.
The distribution of the correlation $C_{i,j}$ between pairs of nodes in B is shown as a function of parameters $n_\text{B}$, $n'_\text{B}$, $n_\text{A}$, $\beta$, and $\sigma$.
Each panel shows the median (solid curve with dots), the range between the minimum and maximum (shaded region), and the $25$th and $75$th percentiles (dashed curves). 
These quantities are all averaged over $10$ network realizations and $10$ initial conditions.
Unless noted otherwise, all parameters are set to the values used in Fig.~\ref{fig3} (indicated by red dots).}\label{fig4}
\end{figure*}
How does IMRS depend on system parameters?
To answer this question, we study the distribution of $C_{i,j}$ (Fig.~\ref{fig4}) and $I_{i,j}$ (Fig.~\ref{fig:fig4-mi} in Sec.~\ref{sec:lambda-C-one-nA-nB-nBp} of Supplemental Material~\cite{sm}) over all $i\neq j\in$~B as functions of parameters $n_\text{B}$, $n'_\text{B}$, $n_\text{A}$, $\beta$, and $\sigma$.
We verify $\lambda_{C_1}<0$ for the entire range of parameter values over which the curves are drawn in Fig.~\ref{fig4} (see Supplemental Material~\cite{sm}, Sec.~\ref{sec:lambda-C-one-nA-nB-nBp} for more details, including parameter dependence of $\lambda_{C_1}$).
As indicated by their $25$th and $75$th percentiles (dashed curves), the cross correlation and mutual information remain low for most node pairs in B for a range of system parameters, with the exception of cases with small $\sigma$.
The medians of these coherence measures are mostly monotonically decreasing functions of $\sigma$ up to the largest value of $\sigma$ ($\approx 1.7$) for which $\lambda_{C_1}<0$ [Fig.~\ref{fig4}(e) and Fig.~\ref{fig:fig4-mi}(e)].
We have $C_{i,j}=1$ at $\sigma=0$, indicating that all nodes in B are perfectly correlated in that case, simply because the isolated oscillators all converge to a common stable fixed point for $\beta=1.5$.
The median cross correlation and the median mutual information appear to be slightly decreasing functions of $n_\text{B}$ and $n'_\text{B}$, while they seem to be approximately constant as functions of $n_\text{A}$ (both for $\sigma=1.5$ and $\sigma=1$) and of $\beta$.
Note, however, that the synchronization stability does depend on $n_\text{A}$: we observe that nodes $1$ and $N$ cannot synchronize stably for $n_\text{A}>6$ for $\sigma=1.5$ [green curves ending at $n_\text{A}=6$ in Fig.~\ref{fig4}(c) and Fig.~\ref{fig:fig4-mi}(c)] but remain stably synchronized up to $n_\text{A}=15$ for $\sigma=1$ [blue curves in Fig.~\ref{fig4}(c) and Fig.~\ref{fig:fig4-mi}(c)].
The loss of synchronization stability for sufficiently large $n_\text{A}$ is likely due to incoherent dynamics of the other nodes in groups A and C (see Supplemental Material~\cite{sm}, Sec.~\ref{sec:dyn-L}). Since these nodes are the only ones that directly influence the dynamics of nodes $1$ and $N$ (and thus their synchronization stability), the larger the number of these dynamically incoherent nodes (i.e., the larger $n_\text{A}$), the more difficult for nodes $1$ and $N$ to stably synchronize.
Overall, we find that IMRS is observed for a wide range of structural and dynamical parameters of the system (see Supplemental Material~\cite{sm}, Sec.~\ref{sec:lorenz-robustness} for similar robustness observed for a continuous-time system).

We also find that the low levels of coherence between node $1$ (or $N$) and the nodes in B are maintained over a range of parameter values, following dependence patterns similar to those of the coherence levels within B (see Supplemental Material~\cite{sm}, Sec.~\ref{sec:cohe-AB}).
Low coherence between periphery and intermediate nodes has also been observed in certain cases of remote synchronization~\cite{Sande2008,Soriano2012} [but with pairs of identically synchronized oscillators in the intermediate part of the network,
which is not compatible with the IMRS condition (iii)].
 
A key aspect of IMRS lies in its behavior against noise.  While the synchronization of nodes $1$ and $N$ is robust against independent noise added to the dynamics in A and C only up to a certain level (which is expected), IMRS is completely insensitive to noise in B, even when the noise level is very high (see Supplemental Material~\cite{sm}, Sec.~\ref{sec:robustness-noise}).
This characteristic robustness of IMRS stems from the mirror symmetry and is also associated with the dynamical incoherence in condition (iii).
In contrast, (remote) synchronization of nodes $1$ and $N$ can be extremely sensitive to noise in B when some nodes in B are identically synchronized.
This is demonstrated using the network topology considered in Ref.~\cite{Sande2008} (see Supplemental Material~\cite{sm}, Sec.~\ref{sec:non-robustness-noise}).

Our demonstration of IMRS challenges the notion that paths of communication between nodes that are exchanging information should be somehow observable.
A particularly striking feature of IMRS we studied here is that the coupling between A and B, as well as B and C, is bidirectional.
This allows information to be transferred from A to C through B, despite the scrambling of that information by the incoherent chaotic dynamics of B, which reduces the amount of shared information in B to a level that is too low for eavesdroppers (as measured by mutual information).
This feature fundamentally sets IMRS apart from a master-slave type of chaos synchronization~\cite{Pecora:1990}, in which the dynamics of B influences that of A and C, but not vice versa, thus prohibiting communication between nodes $1$ and $N$.
Similar master-slave synchronization can be observed even when B is replaced by noise, if the average of the noise is nonzero and its effect is equivalent to parameter change that drives the dynamics into synchrony~\cite{Herzel:1995}.

A defining characteristic of IMRS we demonstrated is the dynamical incoherence within group B, which is allowed in the presence of the mirror symmetry we established as a general condition for observing IMRS.
While we focused on undirected networks here, an analog of mirror symmetry can be formulated for directed networks using the notion of input equivalence~\cite{Stewart:2003}.
Since zero-lag synchronization of distant areas of the brain has been experimentally observed~\cite{Roelfsema:1997,Rodriguez:1999,Soteropoulus:2006}, our results suggest the intriguing possibility that a mirror symmetry is hidden deep inside the synaptic connectivity structure.
We hope that our discovery will spark the interest of many researchers and lead to further discoveries of fundamental connections between hidden network symmetry and emergent collective behavior in complex systems.


This work was supported by the NSFC (Award No.~61274042), CSC, 2013 Doctoral Innovation Funds of SWJTU, and ARO (Award No.~W911NF-15-1-0272).
The authors thank Prof.~Wei Pan for comments on the manuscript.

\clearpage
\onecolumngrid
\setcounter{page}{1}

\setcounter{equation}{0}
\renewcommand{\theequation}{S\arabic{equation}}
\setcounter{figure}{0}
\renewcommand{\thefigure}{S\arabic{figure}}
\setcounter{section}{0}
\renewcommand{\thesection}{S\arabic{section}}

\begin{center}
{\Large\bf Supplemental Material}\\[3mm]
\textit{Incoherence-Mediated Remote Synchronization}\\[1pt] 
Liyue Zhang, Adilson E. Motter, and Takashi Nishikawa
\end{center}

\section{Networks of coupled Lorenz oscillators}
\label{sec:lorenz}

\begin{figure}[h]
\centering
\includegraphics[width=4.5in]{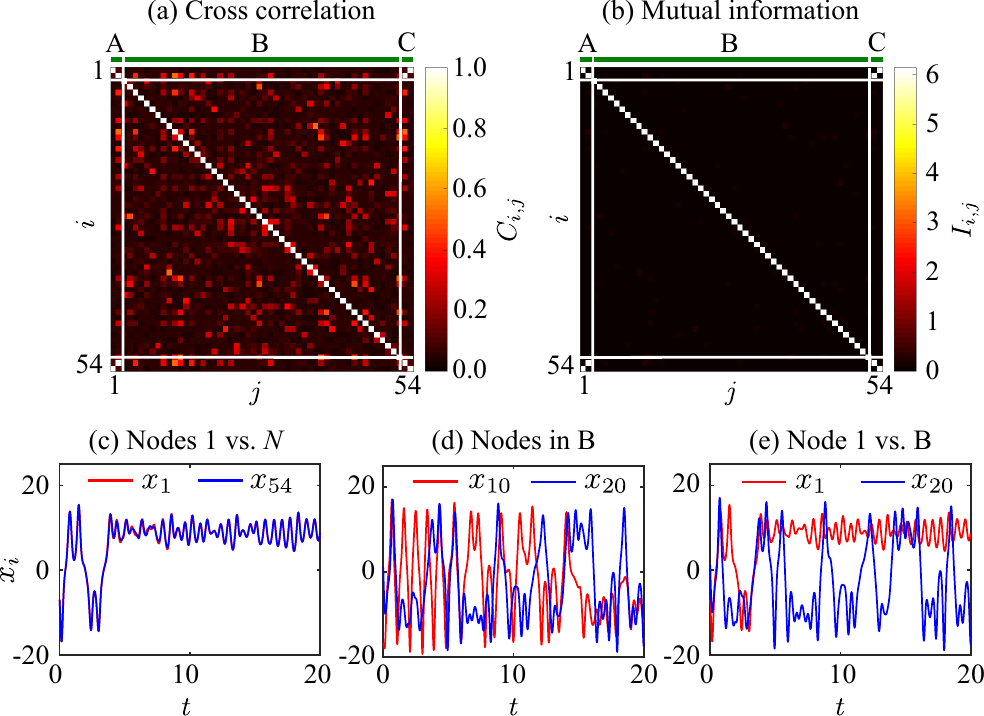}
\caption{
IMRS in a network of $N=54$ coupled Lorenz oscillators described by Eq.~\eqref{eqn:lorenz}.
(a) Pairwise cross correlations $C_{i,j}$.
(b) Pairwise mutual information $I_{i,j}$.
(c)--(e) Oscillator variables $x_i$ as functions of time $t$.
In (c), only the blue curve is clearly visible because the two curves overlap.
}\label{fig8}
\end{figure}

As an example of continuous-time systems exhibiting IMRS, we consider a network of coupled Lorenz oscillators, whose dynamics is governed by
\begin{equation}\label{eqn:lorenz}
\begin{bmatrix}\dot{x}_i\\\dot{y}_i\\\dot{z}_i\end{bmatrix}
=\begin{bmatrix}
\gamma(y_i-x_i)\\
x_i(\rho - z_i)-y_i\\
x_i y_i - \beta z_i
\end{bmatrix}
+\sigma\sum_{j=1}^N A_{ij}\begin{bmatrix}0\\y_i\\0\end{bmatrix},
\end{equation}
which has the same form as Eq.~\eqref{eqn:system} of the main text.
Figure~\ref{fig8} shows results analogous to Fig.~\ref{fig3} of the main text for this system.
We use network topologies constructed by our procedure with $n_\text{B}=50$, $n'_\text{B}=10$, $n_\text{A}=2$, and $p=0.1$ (and thus $N=2n_\text{A}+n_\text{B}=54$).
Figure~\ref{fig8} confirms that, for $\gamma=9$, $\beta=5$, $\rho=19$, and $\sigma=1.25$, we indeed observe IMRS.

\vspace{5mm}

\clearpage
\section{Dependence of IMRS on system parameters}
\label{sec:lambda-C-one-nA-nB-nBp}

\begin{figure*}[h]
\centering
\includegraphics[width=\textwidth]{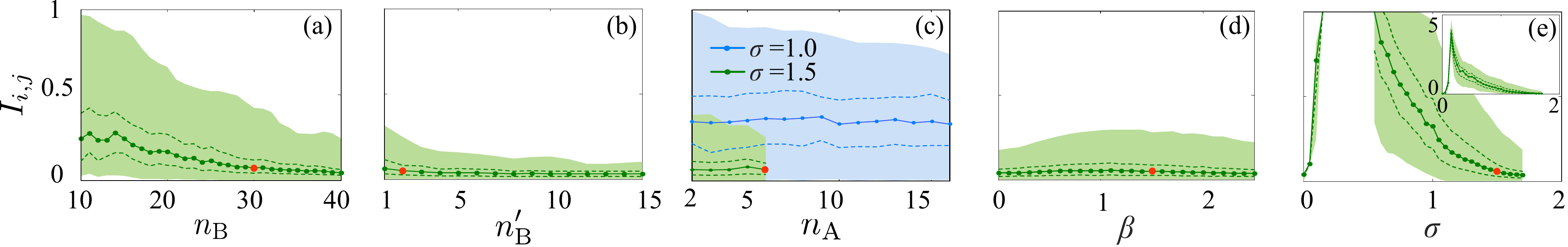}
\caption{
Distribution of mutual information $I_{i,j}$ as a function of system parameters for the system in Eq.~\eqref{discrete} (the counterpart of Fig.~\ref{fig4} for $I_{i,j}$).
In the last panel, the main plot shows only the range $0 \le I_{i,j} \le 1$, while the inset shows the entire plot.
}\label{fig:fig4-mi}
\end{figure*}

\begin{figure*}[h]
\centering
\includegraphics[width=\textwidth]{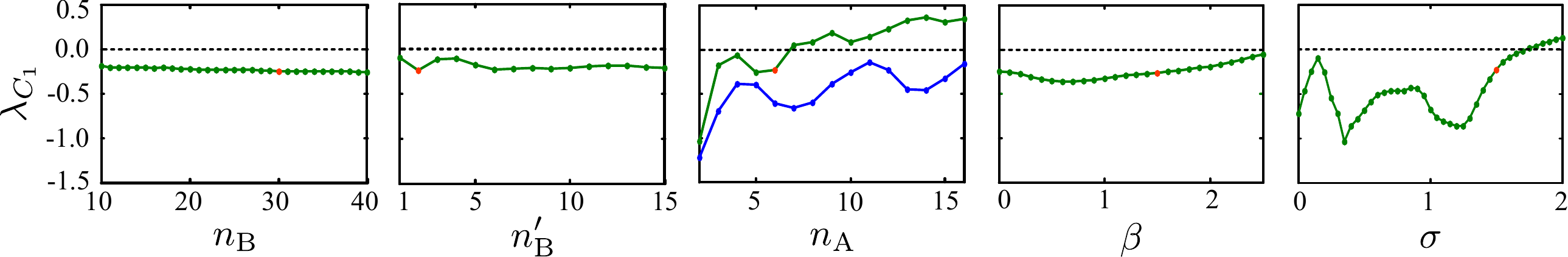}
\caption{
Synchronization stability $\lambda_{C_1}$ between nodes $1$ and $N$ as a function of system parameters for the system in Eq.~\eqref{discrete}.
}\label{fig:lambda-C-one-nA-nB-nBp}
\end{figure*}

To complement the results in Fig.~\ref{fig4} for the system in Eq.~\eqref{discrete}, we also compute the mutual information $I_{i,j}$ and the Lyapunov exponent $\lambda_{C_1}$ for the synchronization of cluster $C_1$ as a function of the parameters $n_\text{B}$, $n'_\text{B}$, $n_\text{A}$, $\beta$, and $\sigma$.
The results are shown in Figs.~\ref{fig:fig4-mi} and \ref{fig:lambda-C-one-nA-nB-nBp}, respectively.
As in Fig.~\ref{fig4}, in each plot the parameters are set to the values used in Fig.~\ref{fig3} (indicated by red dots), except for the one being varied.
Both quantities are averaged over ten network realizations and ten initial conditions.

\vspace{5mm}

\vspace{5mm}

\section{Dynamics of nodes in groups A and C}
\label{sec:dyn-L}

\begin{figure*}[h]
\centering
\includegraphics[width=0.8\textwidth]{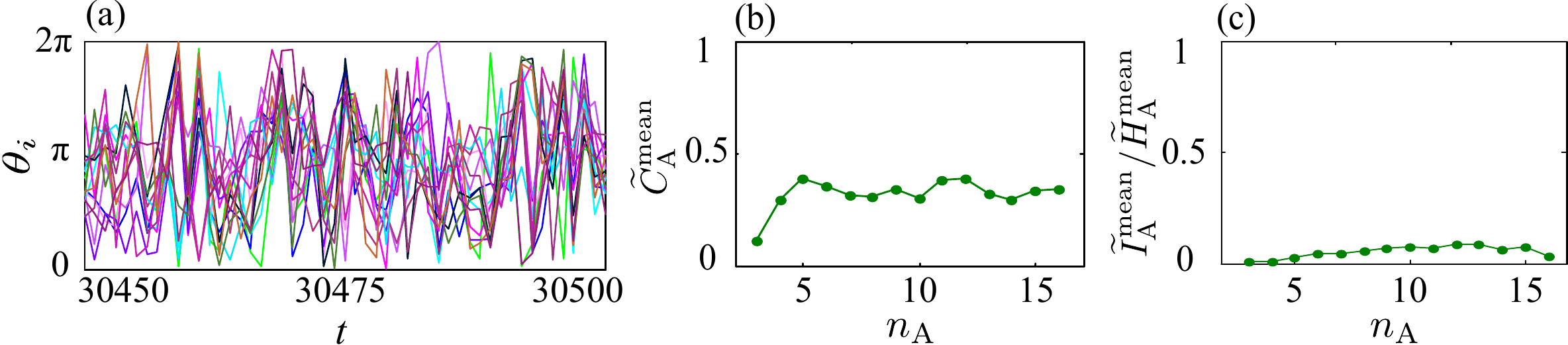}
\caption{
Dynamics of nodes in A other than node $1$ for the system in Eq.~\eqref{discrete}.
(a) Time series for the state variables of these nodes.
(b) Mean cross correlation $\widetilde{C}_\text{A}^\text{mean}$ among the time series in (a) as a function of $n_\text{A}$.
(c) Mean pairwise mutual information $\widetilde{I}_\text{A}^\text{mean}$ among the time series in (a), relative to the mean entropy $\widetilde{H}_\text{A}^\text{mean}$ of these time series, as a function of $n_\text{A}$.
The network structure and parameters are the same as in Fig.~\ref{fig3}.
}\label{fig10}
\end{figure*}

Since each node in group A is identically synchronized with the corresponding node in group C in a system exhibiting IMRS, we only need to consider the dynamics in A.
Figure~\ref{fig10} shows low cross correlation and mutual information observed between the (chaotic) dynamics of the nodes within A (other than node $1$) for the system in Eq.~\eqref{discrete}.
Here we use $\widetilde{C}_\text{A}^\text{mean}$ and $\widetilde{I}_\text{A}^\text{mean}$ to denote the mean value of the correlation $C_{i,j}$ and mutual information $I_{i,j}$, respectively, over all node pairs $i,j\in \text{A}$, $i\neq1$, $j\neq1$.
We normalize $\widetilde{I}_\text{A}^\text{mean}$ by the mean value $\widetilde{H}_\text{A}^\text{mean}$ of the entropy $H_i$ of the time series among the nodes in group A other than node $1$.
We find that the level of coherence is approximately constant as a function of $n_\text{A}$, the number of nodes in A.

\clearpage
\section{Robustness against parameter changes for Lorenz oscillator networks}
\label{sec:lorenz-robustness}

\begin{figure*}[h]
\centering
\includegraphics[width=0.9\textwidth]{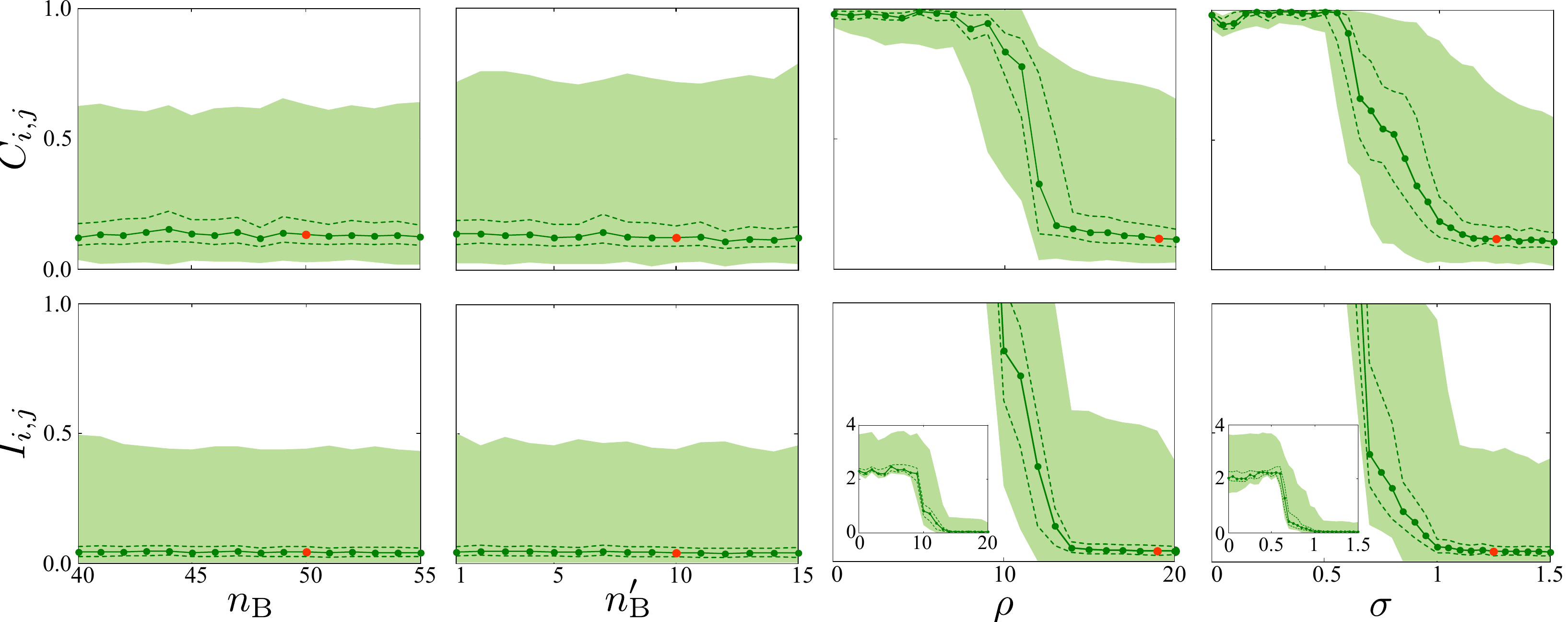}
\caption{
Counterparts of Fig.~\ref{fig4} and Fig.~\ref{fig:fig4-mi} for coupled Lorenz oscillators, showing the influence of system parameters on IMRS.
All parameters are set to the values used in Fig.~\ref{fig8} (indicated by red dots), except for the one being varied.
}\label{fig9}
\end{figure*}

\begin{figure*}[h]
\centering
\includegraphics[width=\textwidth]{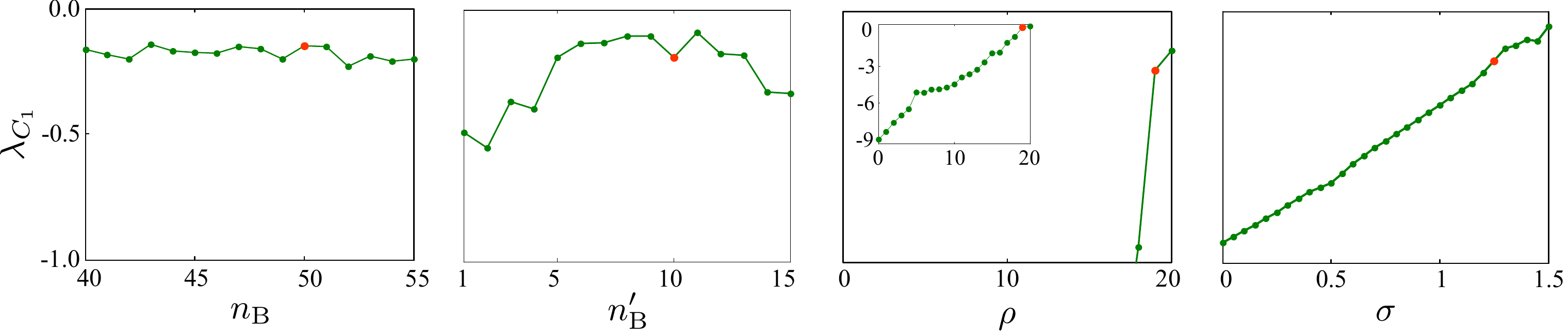}
\caption{
Counterpart of Fig.~\ref{fig:lambda-C-one-nA-nB-nBp} for coupled Lorenz oscillators, showing the synchronization stability $\lambda_{C_1}$ as a function of system parameters.
All parameters are set to the values used in Fig.~\ref{fig8} (indicated by red dots), except for the one being varied.
}\label{fig:lyalorenznew}
\end{figure*}

Here we study the robustness of IMRS for the networks of coupled Lorenz oscillators.
Figure~\ref{fig9} is an analog of Fig.~\ref{fig4} and Fig.~\ref{fig:fig4-mi}, which shows that the cross correlation and mutual information remain low for a range of parameters $n_\text{B}$, $n'_\text{B}$, $\rho$, and $\sigma$.
Similarly, Fig.~\ref{fig:lyalorenznew} is an analog of Fig.~\ref{fig:lambda-C-one-nA-nB-nBp}, which shows the parameter dependence of the synchronization stability between nodes $1$ and $N$.
Note that, for this system, the synchronization is unstable for $n_\text{A}>2$.

\vspace{5mm}

\clearpage
\section{Coherence between node $1$ and group B vs.\ system parameters}
\label{sec:cohe-AB}

\begin{figure*}[h]
\centering
\includegraphics[width=\textwidth]{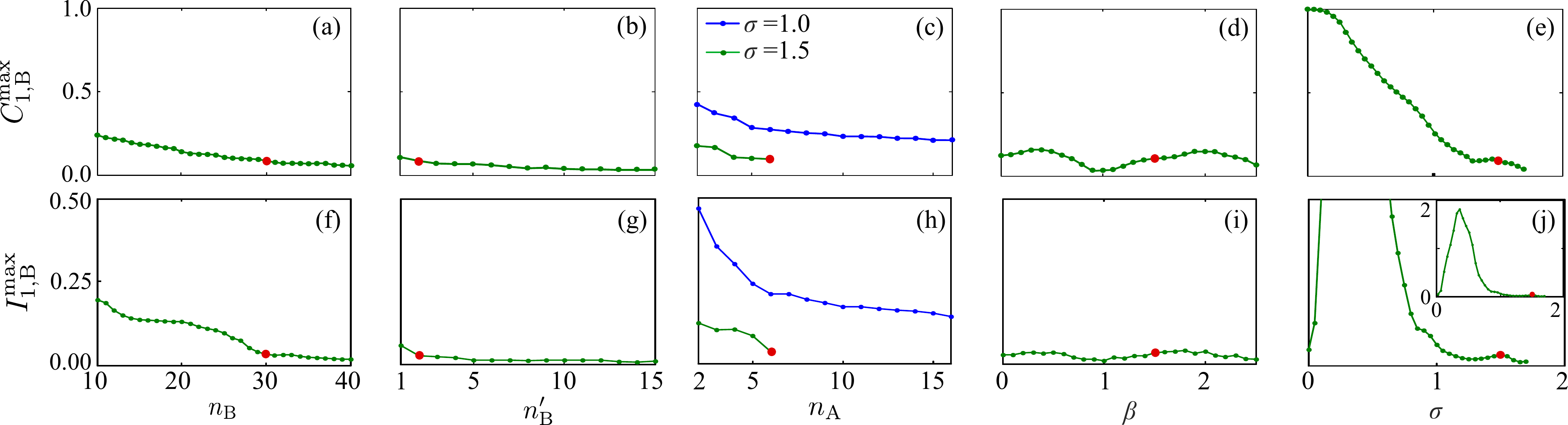}\vspace{-4mm}
\caption{
Influence of system parameters on the coherence between node $1$ and the nodes in B for the system in Eq.~\eqref{discrete}.
Unless noted otherwise, all parameters are set to the values used in Fig.~\ref{fig3} (indicated by red dots).
The top and bottom rows show the cross correlation and mutual information, respectively.
Columns show these quantities as functions of parameters $n_\text{B}$, $n'_\text{B}$, $n_\text{A}$, $\beta$, and $\sigma$.
The averages are taken over ten network realizations and ten initial conditions.
In (j), the main panel shows only part of the plot satisfying $0\le I_{1,\text{B}}^{\max} \le 0.5$, while the inset shows the entire plot. 
}\label{fig:node1-B}
\end{figure*}

Here we measure the coherence between node $1$ and group B by
\begin{equation}\label{eqn:C1Bmax}
C_{1,\text{B}}^{\max}:=\max_{j\in\text{B}}C_{1,j}\quad\text{and}\quad
I_{1,\text{B}}^{\max}:=\max_{j\in\text{B}}I_{1,j},
\end{equation}
which are plotted for the system in Eq.~\eqref{discrete} as functions of parameters $n_\text{B}$, $n'_\text{B}$, $n_\text{A}$, $\sigma$, and $\beta$ in Fig.~\ref{fig:node1-B}.
We see that the behavior of these functions is very similar to Fig.~\ref{fig4} and Fig.~\ref{fig:fig4-mi}.

\clearpage
\section{Robustness against noise}
\label{sec:robustness-noise}

\begin{figure*}[h]
\centering
\includegraphics[width=0.7\textwidth]{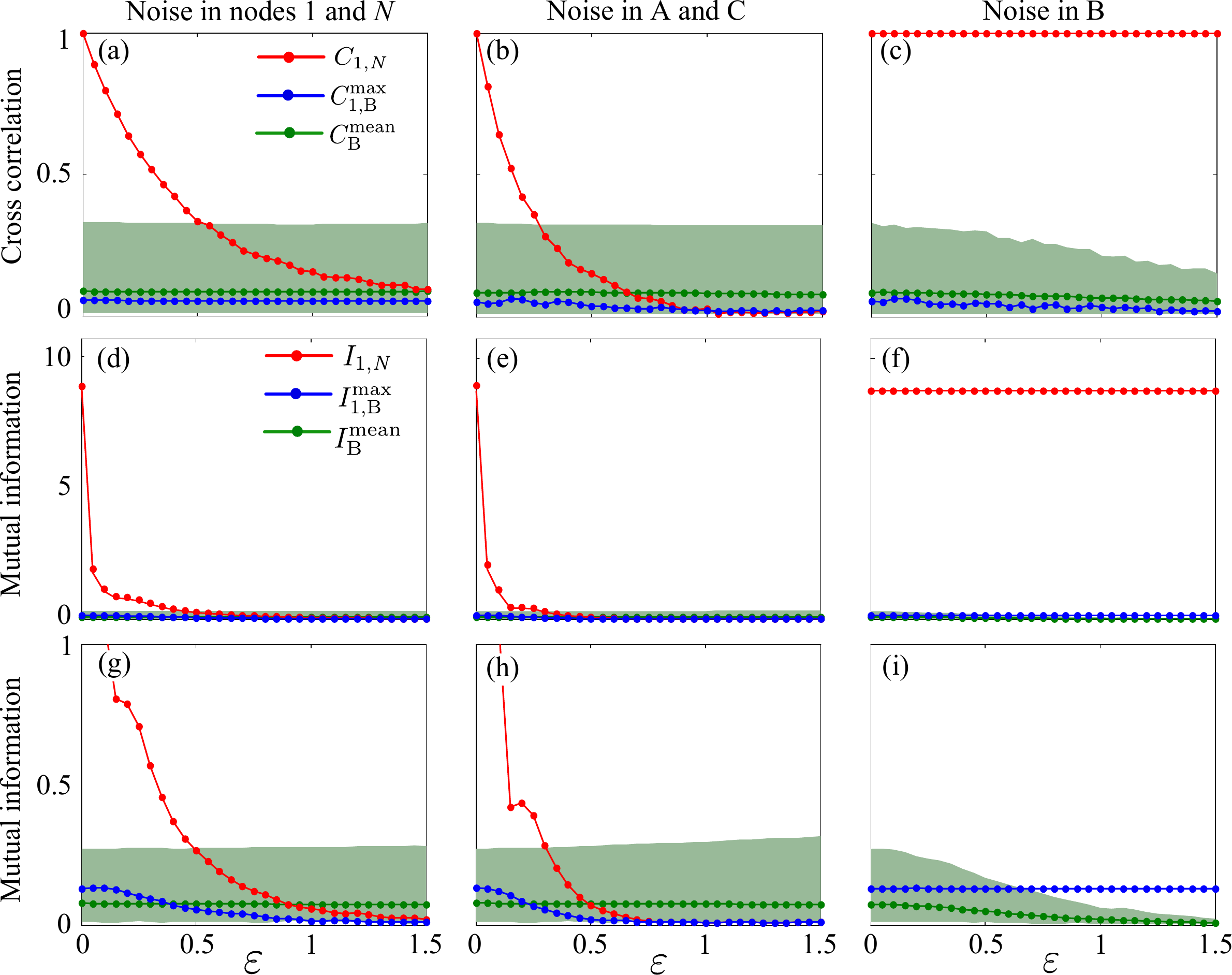}
\caption{
Robustness of IMRS against dynamical noise for the system in Eq.~\eqref{discrete}.
(a)--(c) Cross correlation $C_{1,N}$, $C_{1,\text{B}}^{\max}$, and $C_\text{B}^\text{mean}$ as functions of noise level $\varepsilon$.
(d)--(f) Mutual information $I_{1,N}$, $I_{1,\text{B}}^{\max}$, and $I_\text{B}^\text{mean}$ as functions of $\varepsilon$.
(g)--(i) Vertical magnification of (d)--(f) around zero.
In each panel, the range of $C_{i,j}$ and $I_{i,j}$ over all $i\neq j\in$ B is indicated by green shading.
}\label{fig6}
\end{figure*}

To test the robustness of IMRS against dynamical noise, we add independent Gaussian noise with mean zero and variance $\varepsilon^2$ to the phase variable $\theta_i$ of nodes in different parts of the network described by Eq.~\eqref{discrete}.
We use the parameter values from Fig.~\ref{fig3}.
Regardless of where the noise is added, the coherence among the nodes in B, as well as between node $1$ and the nodes in B, remains low (and approximately constant) as a function of $\varepsilon$.
This can be seen in Fig.~\ref{fig6}, where we plot the quantities 
\begin{equation}\label{eqn:CBmean}
C_\text{B}^\text{mean} := \langle C_{i,j} \rangle_{i\neq j\in\text{B}} \quad\text{and}\quad
I_\text{B}^\text{mean} := \langle I_{i,j} \rangle_{i\neq j\in\text{B}},
\end{equation}
as well as $C_{1,\text{B}}^{\max}$ and $I_{1,\text{B}}^{\max}$ defined in Eq.~\eqref{eqn:C1Bmax}.
The level of synchronization between nodes $1$ and $N$ decreases as $\varepsilon$ increases (as expected) if the noise is added to nodes $1$ and $N$ themselves or to the other nodes in A and C (see $C_{1,N}$ and $I_{1,N}$ in the first and second columns of Fig.~\ref{fig6}).
In contrast, we observe that synchronization is not affected at all by the noise if it is added to B, even for very large $\varepsilon$, despite the fact that nodes $1$ and $N$ are interacting with the nodes in B through network paths connecting them (see $C_{1,N}$ and $I_{1,N}$, the red curves in the third column of Fig.~\ref{fig6}). 

\clearpage
\section{Consequence of having synchronized oscillators in B}
\label{sec:non-robustness-noise}

\begin{figure*}[h]
\centering
\includegraphics[width=0.7\textwidth]{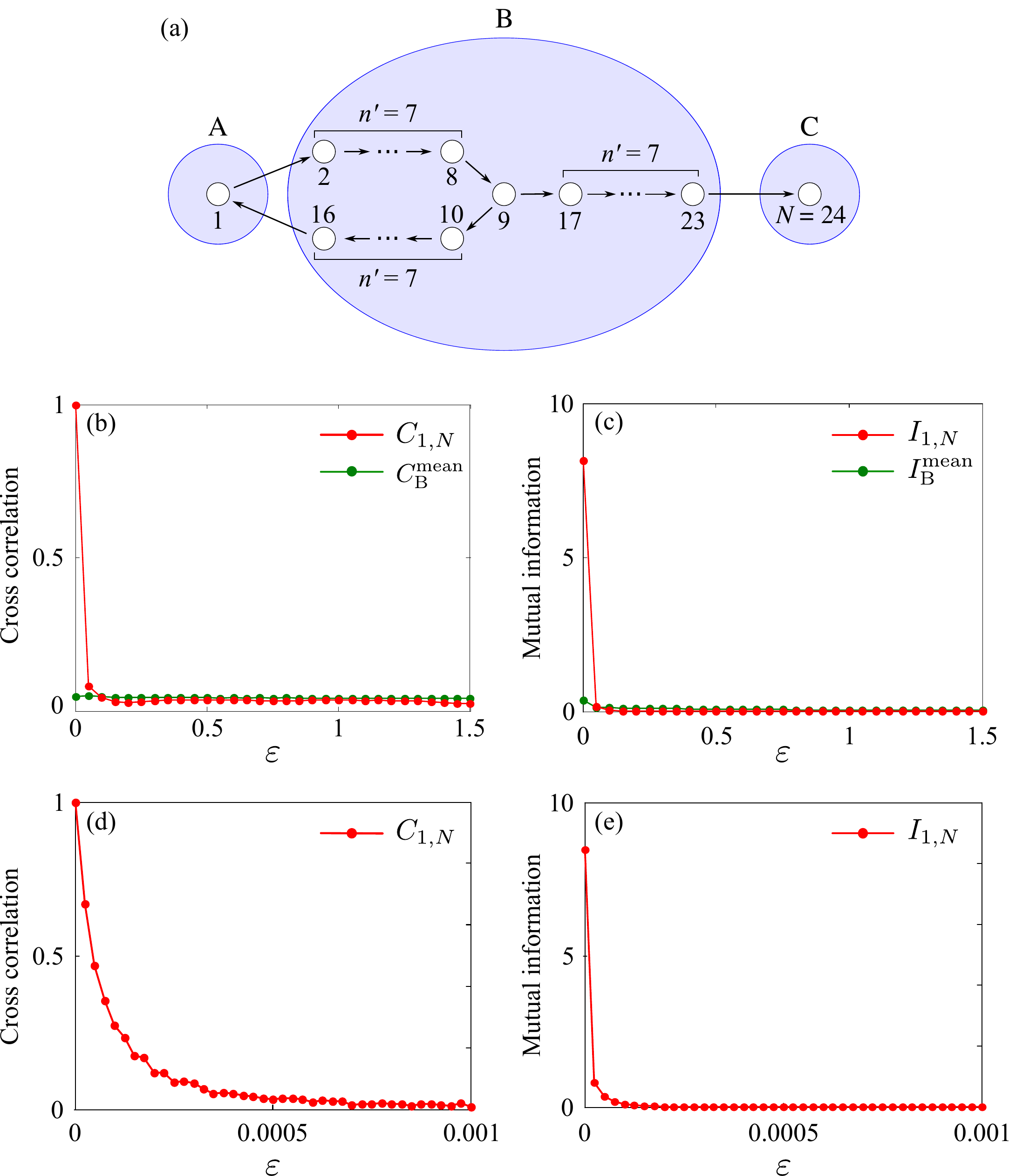}
\vspace{-2mm}
\caption{
Sensitivity of remote synchronization to the addition of independent noise to group B.
(a) Network topology considered.
(b),(c) Cross correlation (b) and mutual information (c) between nodes $1$ and $N$, as well as the average of these quantities among the nodes in B, as functions of the noise intensity $\varepsilon$.
(d),(e) Horizontal magnification of (b) and (c) around zero (the averages not shown).
}\label{fig:B-noise-nonrobust}
\end{figure*}

To demonstrate that remote synchronization can be sensitive to noise in B when some nodes in B are identically synchronized, we consider the (directed) network topology used in the example system of Ref.~\cite{Sande2008}.
This topology is illustrated in Fig.~\ref{fig:B-noise-nonrobust}(a).
Using $n'=7$ for the topology (making it a network of $N=24$ nodes), $\beta=1.5$ for the node dynamics in Eq.~\eqref{discrete}, and $\sigma=10$ for the coupling strength in Eq.~\eqref{discrete}, the system exhibits stable synchronization between nodes $1$ and $N$ [and thus $C_{1,N}=1$ for $\varepsilon=0$, as shown in Fig.~\ref{fig:B-noise-nonrobust}(b)], while most node pairs within B show incoherent dynamics [low $C_\text{B}^\text{mean}$ and $I_\text{B}^\text{mean}$ in Figs.~\ref{fig:B-noise-nonrobust}(b) and \ref{fig:B-noise-nonrobust}(c), respectively, where $C_\text{B}^\text{mean}$ and $I_\text{B}^\text{mean}$ are both defined in Eq.~\eqref{eqn:CBmean}].
However, there are pairs of nodes in B that are identically synchronized in this system~\cite{Sande2008}: nodes $10$ and $17$, nodes $11$ and $18,\ldots,$ as well as nodes $16$ and $23$.
The synchronization of these pairs are sensitive to independent noise [for the same reason we observe sensitivity in the red curves in Figs.~\ref{fig6}(b), \ref{fig6}(e), and \ref{fig6}(h)], which leads to the sensitivity of the synchronization between nodes $1$ and $N$.
This is demonstrated by the sharp drop of the red curves in Figs.~\ref{fig:B-noise-nonrobust}(b)--\ref{fig:B-noise-nonrobust}(e) as $\varepsilon$ is increased from zero, where the phase variable $\theta_i$ of each node in B is subject to independent additive Gaussian noise with mean zero and variance $\varepsilon^2$.
We observe that the sensitivity of the synchronization between nodes $1$ and $N$ is even more drastic than in Figs.~\ref{fig6}(b), \ref{fig6}(e), and \ref{fig6}(h).
We argue that this is due to the amplification of the effect of noise along the path from node $9$ to node $1$ or $N$.
When noise is added to a node in B, the resulting deviation of the node's state propagates to other nodes and is amplified along the path from that node to node $1$ or $N$.
Thus, the synchronization between nodes $1$ and $N$ is more sensitive to noise added to B than to noise added to nodes $1$ and $N$.

\clearpage

\noindent
MATLAB code for generating the random mirror-symmetric networks used in our study (file name: \verb|mirror_sym_net.m|):

\begin{verbatim}
function net = mirror_sym_net(nA,nB,nBp,p)
% Generates a random mirror-symmetric network following the procedure
% described in the paper.
%
% Output: struct variable 'net' with fields, 'nA', 'nB', 'nBp', 'p', 'A'
%
% If no input argument is given, it will use:
%
%   nA = 6, nB = 30, nBp = 2, p = 0.8
%
% The indexing for the adjacency matrix 'net.A' is as follows:
%
%   Group A: i = 1, ... ,nA
%   Group B: i = nA+1, ..., nA+nB
%   Group C: i = nA+nB+1, ..., N (= 2*nA + nB)
%
% Copyright (c) 2017 by Takashi Nishikawa

if nargin == 0
    nA = 6; nB = 30; nBp = 2; p = 0.8;
end
% nL = nA-1;

N = 2*nA + nB;
A = zeros(N);

% Connect nodes in group B.
for i = nA+1:nA+nB
    for j = i+1:nA+nB
        if rand <= p; A(i,j) = 1; A(j,i) = 1; end
    end
end
 
% Connect Node 1 and N to other nodes in groups A and C, respectively. Also
% connect nodes in groups A and C to nodes in group B.
for i = 2:nA
    A(1,i) = 1; A(i,1) = 1;
    A(N, N-i+1) = 1; A(N-i+1, N) = 1;
    nrm=randperm(nB);
    ix = nA + nrm(1:nBp);
    A(ix,i) = 1; A(i,ix) = 1;
    A(ix, N-i+1) = 1; A(N-i+1, ix) = 1;
end

net = struct('nA', nA, 'nB', nB, 'nBp', nBp, 'p', p, 'A', A);

\end{verbatim}

\end{document}